\documentclass[%
 reprint,
 amsmath,amssymb,
 aps,
]{revtex4-2}

\usepackage{graphicx}
\usepackage{dcolumn}
\usepackage{bm}
\usepackage{appendix}

\begin{document}

\preprint{APS/123-QED}

\title{Aluminum Relaxation as the Source of Excess Low Energy Events in Low Threshold Calorimeters}

\author{Roger K. Romani}
 \email{rkromani@berkeley.edu}
\affiliation{
 University of California, Berkeley, \\ Berkeley, California, 94703, United States of America
}

\date{June 7, 2024}

\begin{abstract}
A previously unexplained background called the Low Energy Excess (LEE) has negatively impacted the reach of a variety of low threshold calorimeters including light dark matter direct detection and coherent elastic neutrino-nucleus scattering experiments. The relaxation of stressed aluminum films as mediated by the motion of dislocations may account for these observations.
\end{abstract}

\maketitle


\section{Introduction}

Low threshold calorimeters used for rare event searches (e.g. light dark matter direct detection and coherent elastic neutrino-nucleus scattering) have observed an excess of low energy phonon events often called the Low Energy Excess (or LEE) which has constrained their science reach \cite{EXCESSReview, CRESST2017, CPDV1}. The rate of this background has been observed to rise around $\sim$ 100 eV, with eV scale threshold detectors observing LEE rates approaching the order of Hz. Individual LEE events have been observed to involve no ionization \cite{EDELWEISSHeatOnly}, but to otherwise mimic a particle interaction in pulse shape \cite{CRESSTTimeVariationLEE}. Additionally, this background has been observed to decrease in rate with time since the cryogenic system has been cooled down, and to reset at least in part when the detector is warmed up and re-cooled \cite{CRESSTTimeVariationLEE}. Taken in concert, these observations indicate that the LEE is not a previously unknown particle background, but instead a detector specific effect. Existing models of low energy backgrounds in ionization sensitive detectors \cite{EssigTrackInduced} do not explain either the rate \cite{EssigTrackInduced} or the lack of ionization \cite{EDELWEISSHeatOnly} of LEE events.

Recent work has shown that the relaxation of mechanical stress associated with the detector holding can create events that mimic the LEE in spectral shape and time dependence \cite{StressBackgroundsPaper}. However, even after the detector holding stress was greatly reduced, a residual LEE component remained, which Ref. \cite{StressBackgroundsPaper} attributed to the relaxation of thermally induced stress in the films from which the detector readout sensors were constructed. 

In this paper, I present a model of how the relaxation of aluminum films can create events that are consistent with the LEE. As no first-principles model of LEE events has been proposed, this model represents an important first step toward a testable model of excess events. This model specifically focuses on the relaxation of aluminum, due to its wide use in low temperature devices including TESs \cite{QETPaper} (Transition Edge Sensors), KIDs \cite{AlKID} (Kinetic Inductance Detectors), and superconducting qubit-based devices \cite{QCD, SQUATIdeaPaper}. Different materials with a similar structure (FCC metals with low Peierls stresses, e.g. Au and Cu) may also relax similarly. See Appendix \ref{appendix:other_materials} for further discussion.

This model may also be applicable to observations of relaxing quasiparticle burst rates in quantum circuits \cite{QuasiparticleFreeSeconds, Serniak2018}, but focuses on ``LEE'' observations in calorimeters used as particle detectors.

\section{Overview}

\begin{figure}[b]
\includegraphics[width=0.48\textwidth]{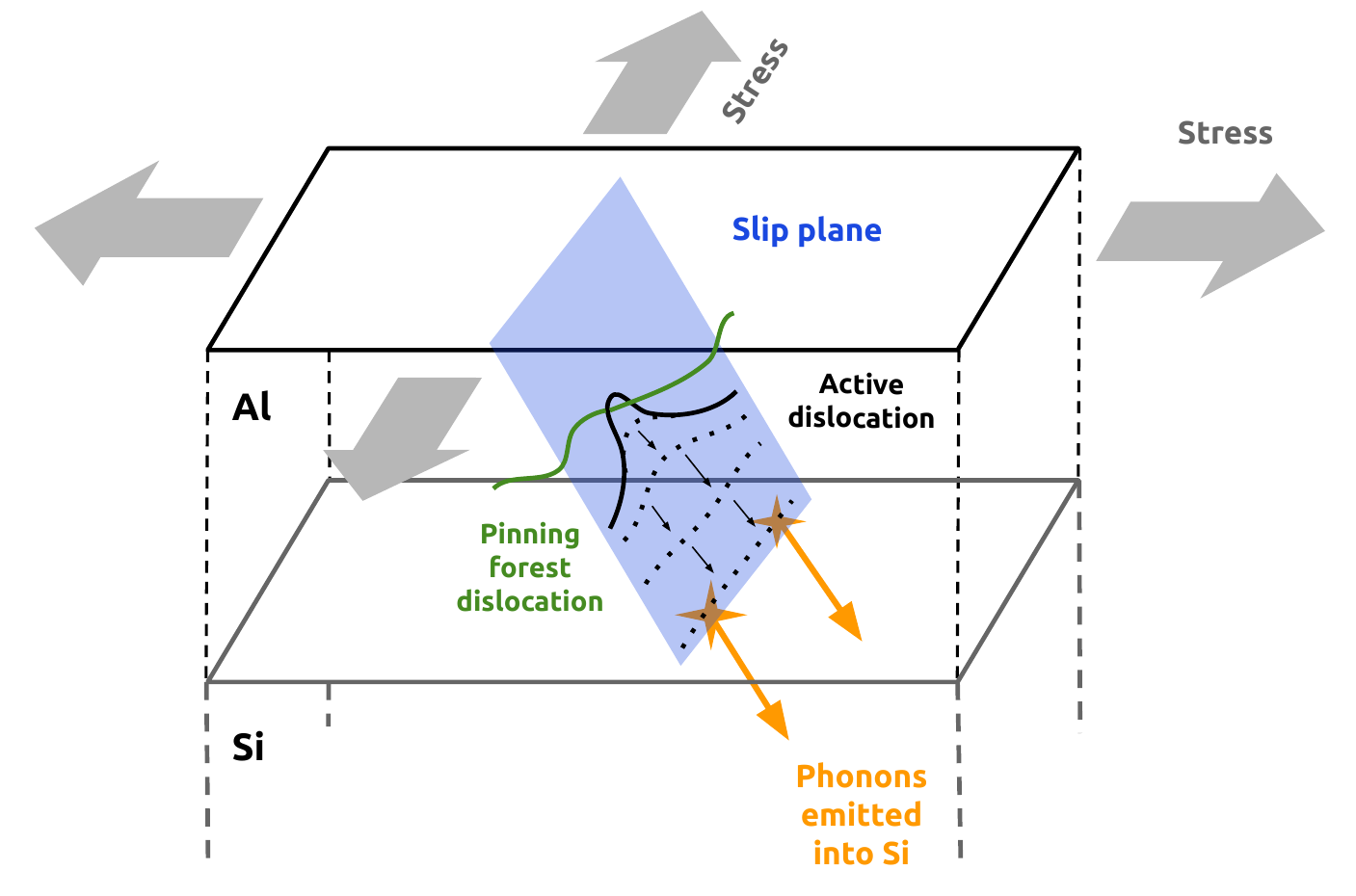}
\caption{\label{fig:diagram_dislocation} A sketch of a the model discussed in this paper. An aluminum film deposited on a silicon (or other crystal) substrate contracts relative to the substrate when cooled to mK temperatures, stressing the film. The relaxation of this film is mediated by dislocations (thick lines), which allow the film to deform through their motion. These dislocations can become stuck or ``pinned'' (shown here pinned against a ``forest dislocation,'' see text), trapping them in a metastable state. These dislocations can tunnel free at later times and accelerate across the film. When these dislocations hit the film-substrate interface, they decelerate quickly, releasing energetic phonons into the silicon substrate.}
\end{figure}

In this paper, I present a model showing that dislocation mediated relaxation in thermally stressed aluminum films produces events that are broadly consistent with the LEE (see Fig. \ref{fig:diagram_dislocation} for a graphical overview of this model).
\begin{itemize}
    \item When deposited on silicon or similar substrates and when cooled to mK temperatures, differential thermal contraction between the substrate and aluminum will induce stresses on the order of 100s of MPa \cite{EkinLowTemp, ThinFilmSim}, well in excess of the aluminum bulk yield strength. 
    \item The stressed film will partially relax through the motion of dislocations \cite{ThinFilmSim}, however, some of these dislocations will become stuck at ``pinning'' sites in a metastable state \cite{HirthDislocations, WorkHardeningFCC, AlMgImpurities}.
    \item These metastable pinned states will later relax through tunneling \cite{MottTunneling}. The relaxation rate of these pinned states scales approximately as $1/t$ for a generic model.
    \item Once dislocations are released from the pinned state, they will accelerate toward the film-crystal interface, where their impact with the film-crystal interface will release energy in the form of phonon bursts. These phonons are emitted in close proximity to the film-crystal interface and are primarily directed down into the crystal, leaving little energy in the film in which the relaxation took place.
\end{itemize}

To model the properties of these dislocation relaxation events, I created two codes (available publicly at Ref. \cite{SimulationGithub}):
\begin{itemize}
    \item A one-dimensional numerical evaluation of the dislocation equations of motion (see Sec. \ref{section:phonon_creation} and Appendix \ref{appendix:simulation}) to simulate the spectrum of phonons emitted from a differential length of dislocation interacting with the film-crystal interface.
    \item A Monte Carlo-based simulation of the spectrum expected from the relaxation of many dislocations given an average dislocation density and stress field (see Sec. \ref{section:spectrum_sim} and Appendix \ref{appendix:simulation}). This simulation uses the results of the previous simulation to inform the efficiency of converting stress energy to above-gap phonons for a given dislocation configuration.
\end{itemize}
These codes rely on linear elasticity theory, and, therefore, fail to accurately model the dynamics of dislocations moving close to the speed of sound or within several interatomic spacings of the film-crystal interface (see Sec. \ref{section:phonon_creation}). In this model, dislocations that create the LEE are common in this regime. Additionally, the Monte Carlo-based simulation neglects the role of grain boundaries, which would serve to impede dislocations, changing both the depinning and phonon emission processes (see Sec. \ref{section:spectrum_sim} for further discussion). See Appendix \ref{sec:future_work} for a discussion of possible future work.

\begin{figure}[b]
\includegraphics[width=0.4\textwidth]{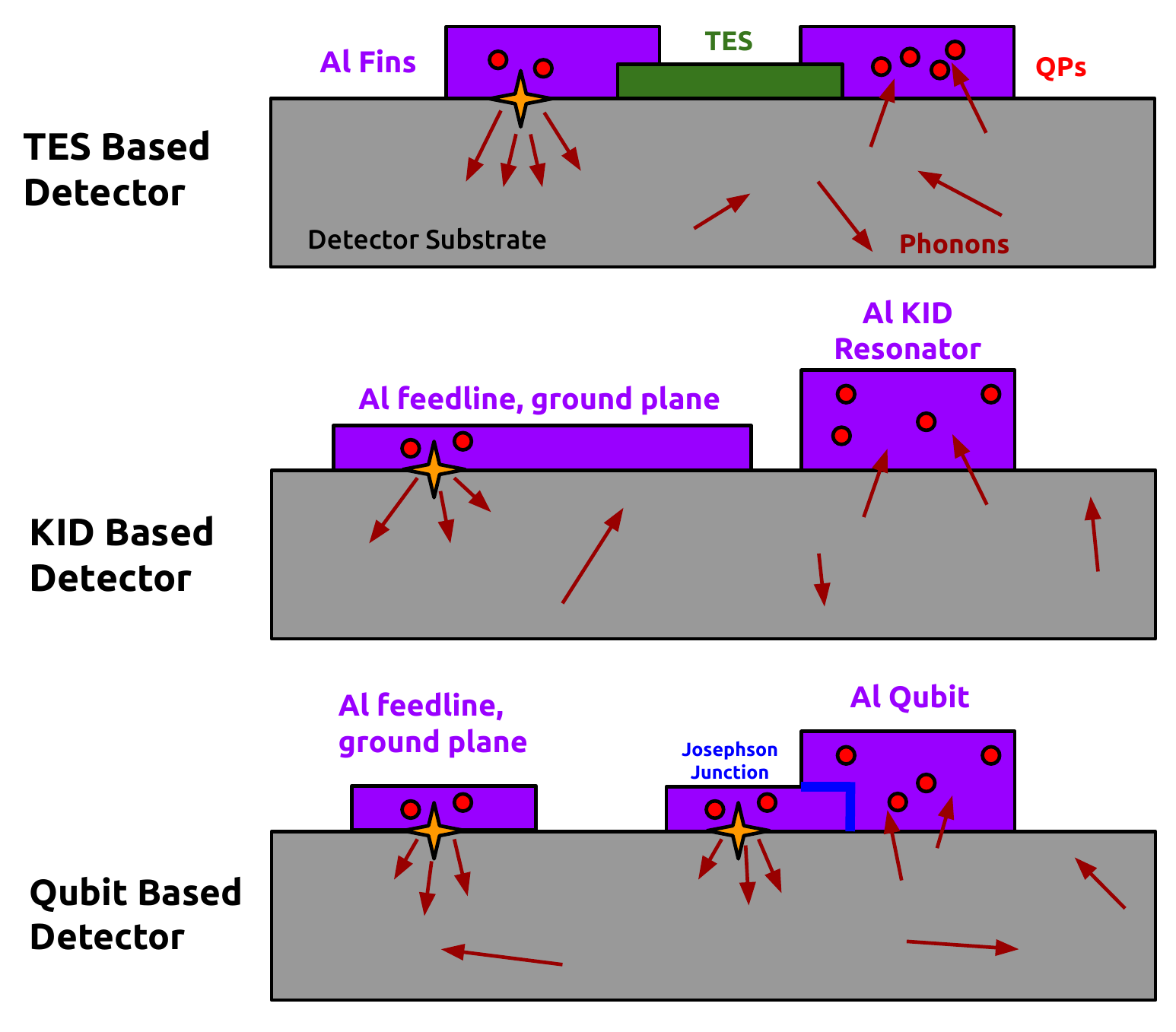}
\caption{\label{fig:diagram_detectors} Sketches of various low temperature detector architectures, showing the creation and sensing of LEE events caused by aluminum relaxation. In each detector type (TES, KID, qubits e.g. \cite{QCD, SQUATIdeaPaper}), aluminum (purple) is commonly used in constructing phonon collection fins, wiring, resonators, and Josephson junctions. This model predicts that aluminum relaxation (orange stars) creates bursts of phonons (dark red) in the detector substrates, which in turn create quasiparticles (red) in the detector superconductors. Depending on the rate and energy scale, these can appear as individual events, sub-threshold noise, or a residual quasiparticle population (``quasiparticle poisoning''). Relaxation may also inject quasiparticles directly into the detector superconductors (see text). Sketches are not to scale.}
\end{figure}

I will conclude by summarizing experimental evidence for this model, highlighting areas of tension between my simulations and experimental results, and by discussing the implications for low threshold cryogenic calorimeter design and operation (see Fig. \ref{fig:diagram_detectors}).

\section{Metal Film Stress and Deformation}
\label{section:film_stress_deformation}

When thin aluminum films (used to construct e.g. phonon collection fins or wiring in cryogenic DM detectors) are deposited on common crystalline substrates (e.g. silicon, sapphire and germanium) and are cooled to mK temperatures, they will contract relative to the substrate by a factor of $\epsilon \approx 4 \times 10^{-3}$ \cite{EkinLowTemp}. This biaxial strain induces a stress of $\sigma \approx $ 400 MPa, well in excess of both the $\sim 1 - 10$ MPa Peierls stress \cite{AlPeierlsStress} (inherent to dislocation transport through the crystal lattice) and 100 - 200 MPa yield stresses measured in thin aluminum films \cite{ThinAlFilmDeformation}.

Metals deform through the motion of dislocations, linelike defects in the crystal structure which are driven by stress fields. To completely relax a $\sim 4 \times 10^{-3}$ strain, dislocations at a density of  $\sim 10^{10}$ cm$^{-2}$ = (100 nm)$^{-2}$ would need to travel $\sim$ 100 nm \cite{HirthDislocations}. These scales indicate that dislocations will travel distances on the order of the dislocation spacing and a significant fraction of the film thickness. Simulations of this process confirm these estimates, and suggest that the residual stresses are on the order of 100 MPa \cite{ThinFilmSim}.

\section{Dislocation Pinning and Depinning}
\label{section:depinning}

As they travel through the metal crystal, individual dislocations may become stuck on obstacles (often called ``pinning sites'' or ``locks''), becoming trapped in metastable excited states. Both dislocations that are not coplanar with the moving dislocation (``forest dislocations'') \cite{DislocationLockSim} and substitutional impurity atoms \cite{AlMgImpurities} are plausible candidates for these pinning sites. A detailed comparison of these two pinning mechanisms will be left for future work, and will not be considered further here, where I will focus on forest dislocation pinning.

In either case, a dislocation lock will immediately fail when the resolved shear stress on the dislocation exceeds a critical stress
\begin{eqnarray} \label{eqn:stress_length_relationship}
    \tau_c = \alpha \mu b/L \approx \alpha \frac{(6 \mathrm{GPa})}{L / (1 \mathrm{nm})}
\end{eqnarray}
where $\mu$ is the shear modulus, $b$ is the Burgers vector, and $L$ is the length of the dislocation pulling on the pinning site (longer dislocations will induce more force on a pointlike pinning site) \cite{AlMgImpurities}. $\alpha$ is a dimensionless constant ranging from $\sim 0.8$ \cite{DislocationLockSim} in the case of certain forest dislocation locks (Lomer-Cottrell locks) to $\sim 0.01$ in the case of certain substitutional atom pinning sites \cite{AlMgImpurities} which quantifies the strength of the pinning.

Dislocations with an actual resolved shear stress $\tau$ on the pinned dislocation close to the critical stress $\tau_c$ can tunnel from the metastable pinned state into the free state at arbitrarily low temperatures. Mott \cite{MottTunneling} estimates the rate of this tunneling as
\begin{eqnarray}
    r(\tau) = \nu_D \exp \bigg(-2 \mathbb{A} a \eta(\tau) \sqrt{\frac{2 M W_0}{\hbar^2}} \bigg)
\end{eqnarray}
where $\eta = (1 - \tau / \tau_c)$, $\nu_D$ is the Debye frequency, $\mathbb{A}$ is a constant Mott takes to be unity, $a$ is the interatomic spacing, $W_0$ is an activation energy which Mott takes as $\sim 1$ eV, and $M$ is the mass per unit width of the tunneling entity which I take as the mass of one aluminum atom (although other estimates imply it could be somewhat lower \cite{CopperTunnelingMass}). Using these estimates, the tunneling rate becomes
\begin{eqnarray}
    r(\tau) \approx \nu_D \exp \big( -900 \eta(\tau) \big) = \nu_D \exp ( - \Gamma \eta(\tau))
\end{eqnarray}
Note that different authors give significantly different estimates of these factors, see e.g. Refs. \cite{DislocationTunneling} and \cite{ CopperTunnelingMass}.

The total depinning rate $R(t)$ (where $t$ is the time after cooldown) for a large number $N$ of pinned dislocations in a metal film can be estimated by assuming the probability distribution of $\eta$ is flat and ranges between 0 and 1. This is equivalent to the assumption that $\tau$ is randomly distributed from 0 to $\tau_c$ at every pinning site, which appears reasonable given the pre-stress dislocation starting position (setting $\tau$), pinned dislocation length (setting $\tau_c$), and local stress field (set by e.g. the location of close by grain boundaries) would be expected to vary quasi-randomly for each individual pinned dislocation. In reality, experiments are only sensitive to $\eta$ in a narrow range around $\eta \approx 0.05$, given the need to relax on time scales comparable to the length of an experiment. In any case, taking the ``flat $\eta$'' approximation, I integrate over $\eta$ to obtain
\begin{eqnarray}
    R(t) = N \int_0^1 r(\eta) \exp \big( - r(\eta) t \big) d \eta  \\
    = \frac{N}{\Gamma t} \Big( \exp(-\nu_D t e^{- \Gamma}) - \exp(-\nu_D t)  \Big) 
\end{eqnarray}
where the factor in parenthesis approaches one for $\Gamma >> \ln(\nu_D t)$ and $\nu_D t >> 1$, leaving us with
\begin{eqnarray} \label{equation:rate}
    R(t) \approx \frac{N}{\Gamma t}
\end{eqnarray}

Critically, this estimate indicates that the upset rate falls off with $1/t$ rather than exponentially. The effective measured time constant of the LEE rate will, therefore, scale with the time at which the measurement was taken (consistent with observations both in detectors \cite{CRESSTTimeVariationLEE} and in quantum circuits \cite{QuasiparticleFreeSeconds}). The $1/t$ scaling only holds assuming that the probability distribution of $\eta$ is relatively flat; subsequent warm up/cooldown cycles (in, e.g., Ref. \cite{CRESSTTimeVariationLEE}) may alter this scaling.

Detailed atomistic simulations of the dislocation depinning process are needed to validate this model, and to improve depinning rate estimates (i.e., to model $\Gamma$).

\section{Phonon Creation}
\label{section:phonon_creation}

After unpinning, dislocations will accelerate through the film with essentially no friction. Both electron and phonon scattering is frozen out due to the superconductivity of the film and the low temperature, and the radiation of energy from the dislocation through interactions with the Peierls potential is generally agreed to be small \cite{DislocationDragging, DislocationInertia}. Eventually, these dislocations will hit an obstacle, such as the film-substrate interface, grain boundaries, or forest dislocations. In general, the deceleration of these dislocations will be damped through the radiation of phonons \cite{HirthDislocations} . 

In this section, I will sketch a classical version of the deceleration process following Ref. \cite{HirthDislocations}. I will only explicitly consider dislocation interactions with the film-substrate interface, as this interface should remain essentially unchanged throughout the dislocation interaction. It is important to emphasize that this classical continuum approach undoubtedly neglects important effects, especially the damping and phonon emission from the dislocation within several $b$ of the interface and around $c_t$. Better simulation of this process (possibly quantizing the phonon emission process as in Ref. \cite{Dislon}) will be left to future work.

The force per unit length of a dislocation near the film-crystal interface can be calculated with the ``image dislocation'' method \cite{IntefaceForce} as
\begin{eqnarray}
    F^* = \frac{dF}{dL} = \tau b + \frac{\mu_f b^2}{2 \pi} \frac{\mu_s - \mu_f}{\mu_s + \mu_f} \frac{1}{x}
\end{eqnarray}
where $x$ is the distance between the dislocation and film-substrate interface along the slip plane, $\mu_f$ and $\mu_s$ are the shear moduli of the film and substrate respectively, and $\tau$ is the resolved shear stress on the dislocation due to bulk stress in the film. Reference \cite{HirthDislocations} gives the effective mass per unit length of the dislocation as
\begin{eqnarray}
    m^* = \frac{\mu_f b^2}{4 \pi c_t^2} \ln \bigg( \frac{c_t}{\gamma \omega b} \bigg)
\end{eqnarray}
such that at every given point, the effective vibration frequency of the dislocation (and of emitted phonons) is given by Ref. \cite{HirthDislocations} as
\begin{eqnarray}
    \omega(x) = \sqrt{\frac{\frac{dF^*}{dx}}{m^*}} = \frac{c_t}{x}\sqrt{2 \frac{\mu_s - \mu_f}{\mu_s + \mu_f} \frac{1}{\ln \big( c_t/(\gamma \omega (x) b) \big)}}
\end{eqnarray}
where $\gamma \approx$ 1.78 and $c_t$ is the transverse speed of sound in the film.

The damping force on a vibrating dislocation \cite{HirthDislocations} is 
\begin{eqnarray}
    F_\mathrm{rad}^* = \frac{d F_{\mathrm{rad}}}{dL} = \frac{\mu_f b^2 \omega}{8 c_t^2} v
\end{eqnarray}
In addition to this damping term, the dislocation loses energy through reductions in its effective mass $m^*$, such that the total energy radiated into the phonon system by a differential length of dislocation is
\begin{eqnarray}
    \frac{\partial E_{\mathrm{rad}}}{\partial L} = \int \bigg( F^*_{\mathrm{rad}} v + \frac{1}{2} v^2 \frac{d m^*}{dt} \bigg) dt
\end{eqnarray}
The energy radiated into above-gap phonons can be calculated by only summing energy emitted by the phonon system while $\hbar \omega > 2 \Delta_{Al}$. In general, these phonons will cause events in detectors, given their ability to create the quasiparticles which detection techniques are sensitive to. TES-based detectors have some additional sensitivity to sub-aluminum-gap phonons, given that phonons can be directly absorbed in the TES itself.

\begin{figure}[b]
\includegraphics[width=0.5\textwidth]{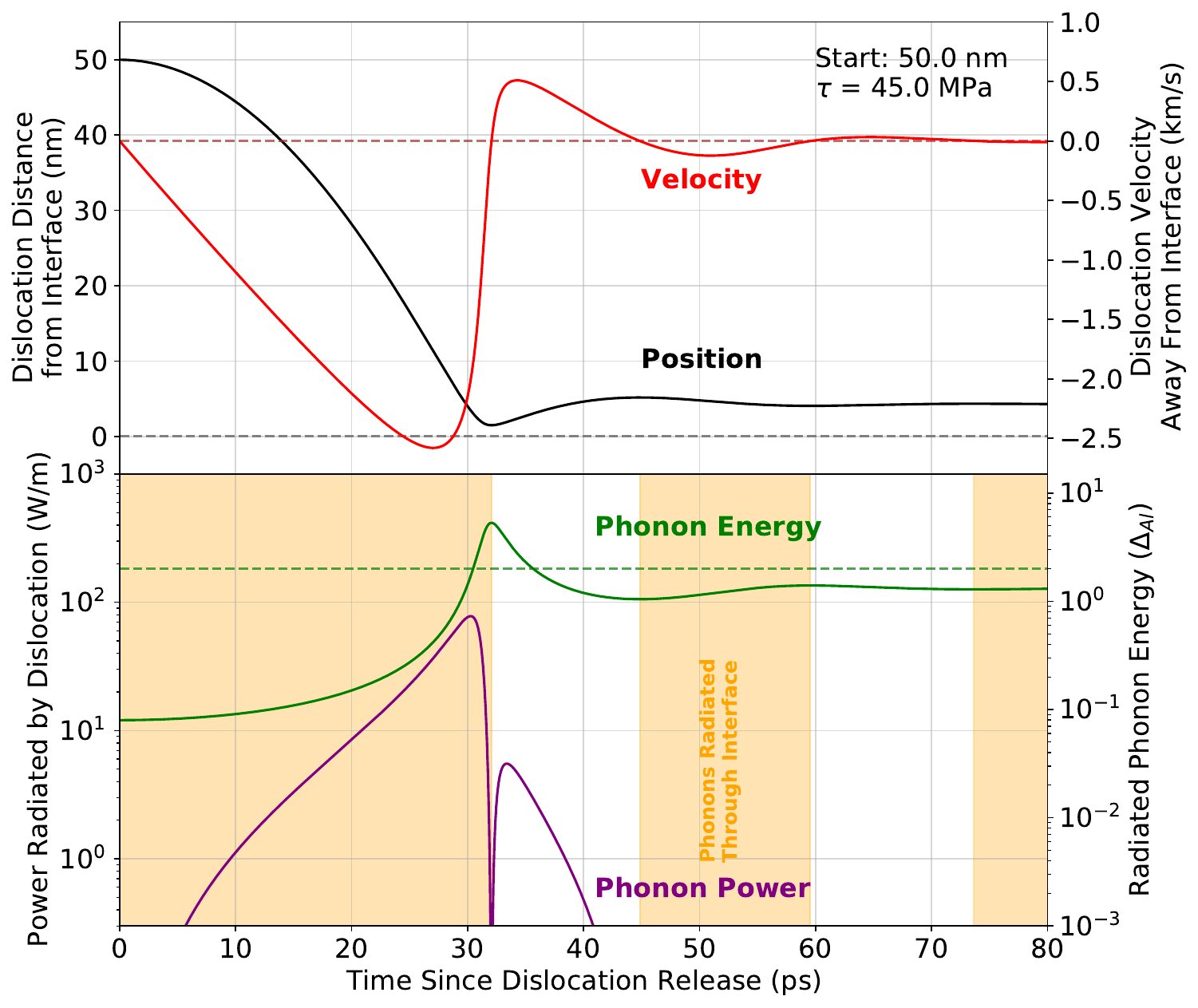}
\caption{\label{fig:dislocation_motion} The simulated motion and phonon radiation of a typical differential length of dislocation (starting 50 nm away from the film-substrate interface, in a 45 MPa resolved shear stress field). Regions where the dislocation is emitting phonons into the substrate are shaded in orange. The green dotted line is at energy 2$\Delta_{Al}$; phonons radiated with an energy above this line (and possibly below, see text) will be measured by detector readout sensors.}
\end{figure}

Using these equations of motion, the deceleration of the dislocation and the properties of associated phonon radiation can be calculated numerically (see Fig. \ref{fig:dislocation_motion}, Appendix \ref{appendix:simulation}).

\begin{figure}[b]
\includegraphics[width=0.5\textwidth]{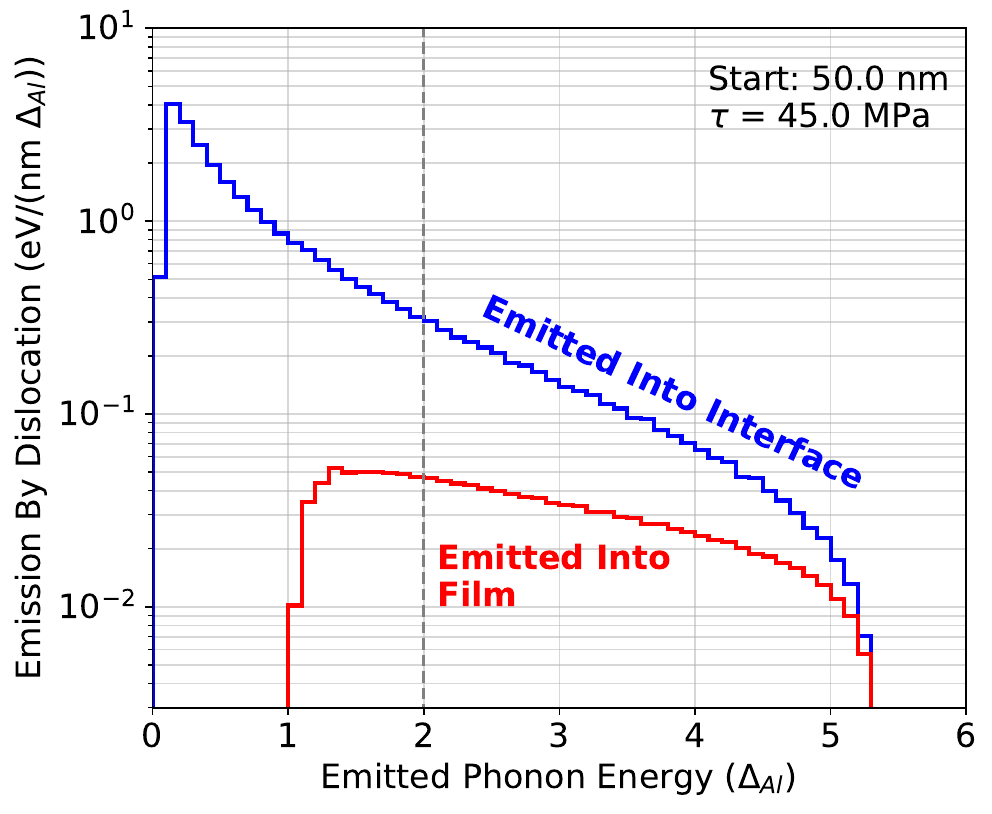}
\caption{\label{fig:dislocation_emission_spectrum} The simulated spectrum of phonons emitted from the dislocation differential length shown in Fig. \ref{fig:dislocation_motion}. The blue spectrum shows the spectrum of phonons emitted downward into the film-substrate interface, and red shows the spectrum of phonons emitted upwards into the film. The dotted gray line is at $2\Delta_{Al}$.}
\end{figure}

\begin{figure}[b]
\includegraphics[width=0.46\textwidth]{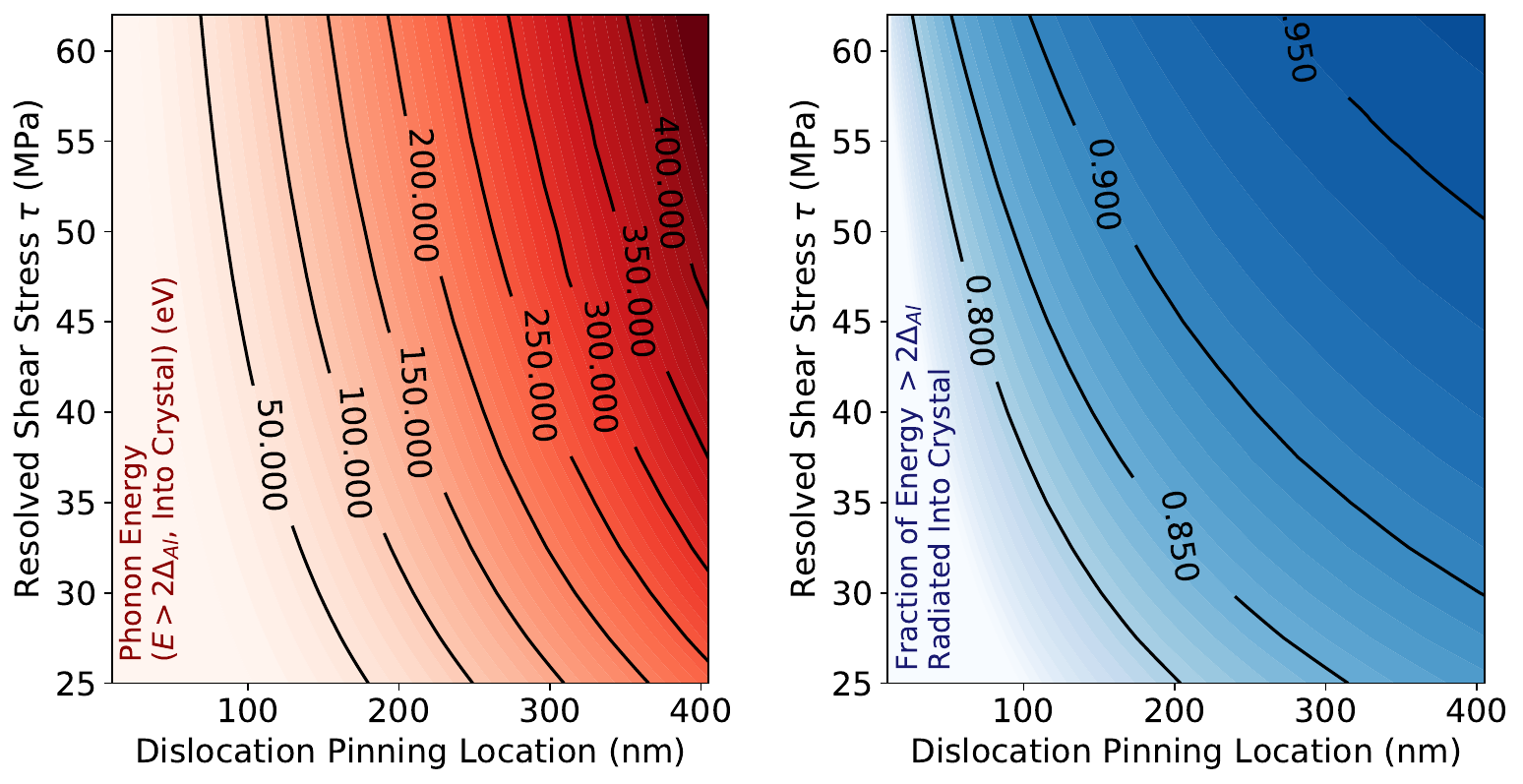}
\caption{\label{fig:dislocation_energy_2d} The simulated energy of above-aluminum-gap phonon energy radiated into the crystal substrate through the film-substrate interface (Left), and the fraction of emitted above-aluminum-gap phonons which are directed into the interface, as opposed to into the film (Right). For clarity, contours on the right plot are cut off at 0.8. Note that for initial pinning locations beyond $\sim$ 65 nm, the dislocation becomes supersonic.}
\end{figure}

Imposing phonon-dislocation momentum conservation (as in Ref. \cite{CoffeyDislocationRadiation}) implies that phonons which cause the dislocation to decelerate are directed through the crystal-film interface at close range ($\sim$nm). Simulations show that $\sim$80-90\% of the above-gap phonon energy is radiated into the crystal, with only a small fraction of above-gap energy radiated backwards into the film (see Figs. \ref{fig:dislocation_emission_spectrum}, \ref{fig:dislocation_energy_2d}).

Several effects likely serve to increase the amount of energy deposited into the phonon system (vs. absorbed locally in the film) compared to the behavior simulated here. First, the phonon damping given by Ref. \cite{HirthDislocations} was calculated for dislocations with $v << c_t$. This linear first-order approximation of the phonon damping likely under-estimates the amount of damping the fast-moving dislocation experiences while first interacting with the interface, implying that there is more emission into the crystal than simulated here. Second, any phonon emitted into the film has a high probability of leaking energy into the crystal, either through escaping without interacting (given the $\sim \mu$m mean free paths of close-to-gap phonons \cite{MartinisSavingQubits}) or through downconversion in which quasi-isotropic secondary phonons are emitted. Finally, this simulation assumes detectors are only sensitive to above-aluminum-gap phonons, which neglects the often considerable amount of sub-gap phonon sensitivity in TES-based detectors. Vastly more sub-gap phonons are emitted into the crystal phonon system compared to into the film (see Fig. \ref{fig:dislocation_emission_spectrum}), meaning that detectors sensitive to these phonons are expected to see less energy absorbed locally.

Broadly, these events should appear to be approximately consistent with substrate phonon events (e.g. calibration events and DM interactions) in which all energy partitions equally between all readout channels observing the same crystal (see, e.g., \cite{CRESSTSingles}). Due to the sub-ns time scales involved, they should also deposit energy quickly compared to the time scales of the sensor.

Improved simulations (see Appendix \ref{sec:future_work}) would greatly aid in more accurately modeling the dislocation deceleration and phonon emission process, and lead to more realistic estimates of the amount of energy absorbed in the film as opposed to emitted into the crystal phonon system. Quantizing the phonon emission process (using e.g. the dislon approach \cite{Dislon}) may also be important to more accurately simulating this model. As discussed above, these improvements would be expected to bring the fraction of phonons emitted into the crystal vs. the film into closer agreement with experimental data.

It is more difficult to estimate the effect of improving simulations on the reconstructed above-gap phonon energy emitted into the crystal during dislocation deceleration (and, therefore, the energy scale of relaxation events). Higher order corrections to the phonon emission process would be expected to increase the rate of deceleration close to the interface, shifting the phonon emission spectrum to higher energies. As an absolute upper bound, these simulations indicate that around 15\% of the initial stored dislocation energy is converted to above-gap phonon energy in the crystal, meaning that if dislocation energy is converted into above-gap phonons with 100\% efficiency, no more than an order of magnitude shift in event energy is expected. While such a large shift seems implausible, the effects on the spectrum of events created by this process (see Fig. \ref{fig:simulated_spectrum}) could be compensated for by assuming that the dislocation mean free path is somewhat smaller than expected (e.g. a higher dislocation density than expected).

\section{Spectrum Simulation}
\label{section:spectrum_sim}

In general, the phonon energy released by a dislocation relaxation event depends on the resolved shear stress experienced by the dislocation and the distance between the dislocation pinning point and the interface.

To simulate the spectrum of energies that would be expected for dislocation relaxation events of this type, I sampled the distribution of possible dislocation resolved shear stresses and starting locations. Given a dislocation density of $\rho_D$, the distance that an otherwise unobstructed dislocation was pinned away from the film interface was drawn from the distribution
\begin{eqnarray}
    \mathrm{Pr}(h_{\mathrm{max}})_\tau = \rho_D L(\tau) e^{-h_{\mathrm{max}} / (\rho_D L(\tau))}
\end{eqnarray}
where $L(\tau)$ is the length of the pinned dislocation (see Eqn. \ref{eqn:stress_length_relationship}). Loosely taking the results in Ref. \cite{ThinFilmSim} (which were performed in two rather than three dimensions and assume a manually set dislocation source density of 50 $\mu\mathrm{m}^{-2}$), I simulate three spectra for $\rho_D$ = (30 nm)$^{-2}$, (50 nm)$^{-2}$, and (100 nm)$^{-2}$. These dislocation densities are broadly consistent with the wide range of values reported in the literature (see, for example, Refs. \cite{DislocationDensity1, AlInternalFriction, DislocationDensity3}). I  neglect the role of grain boundaries, which for small grains may shorten either the length or mean free path of dislocations. 

Resolved shear stresses $\tau$ were determined by assuming that the tensile stress $\sigma_0$ was constant across from the film, and by sampling over the orientations of the slip plane relative to the applied tensile stress, parameterized by the azimuthal angle $\theta$.
\begin{eqnarray}
    \mathrm{Pr}(\theta) = \cos (\theta) \\
    \tau(\theta) = \frac{\sigma_0}{2} \sin (2 \theta)
\end{eqnarray}
Again referencing the results of the simulation in Ref. \cite{ThinFilmSim}, I take $\sigma_0 = 120$ MPa.

\begin{figure}[b]
\includegraphics[width=0.48\textwidth]{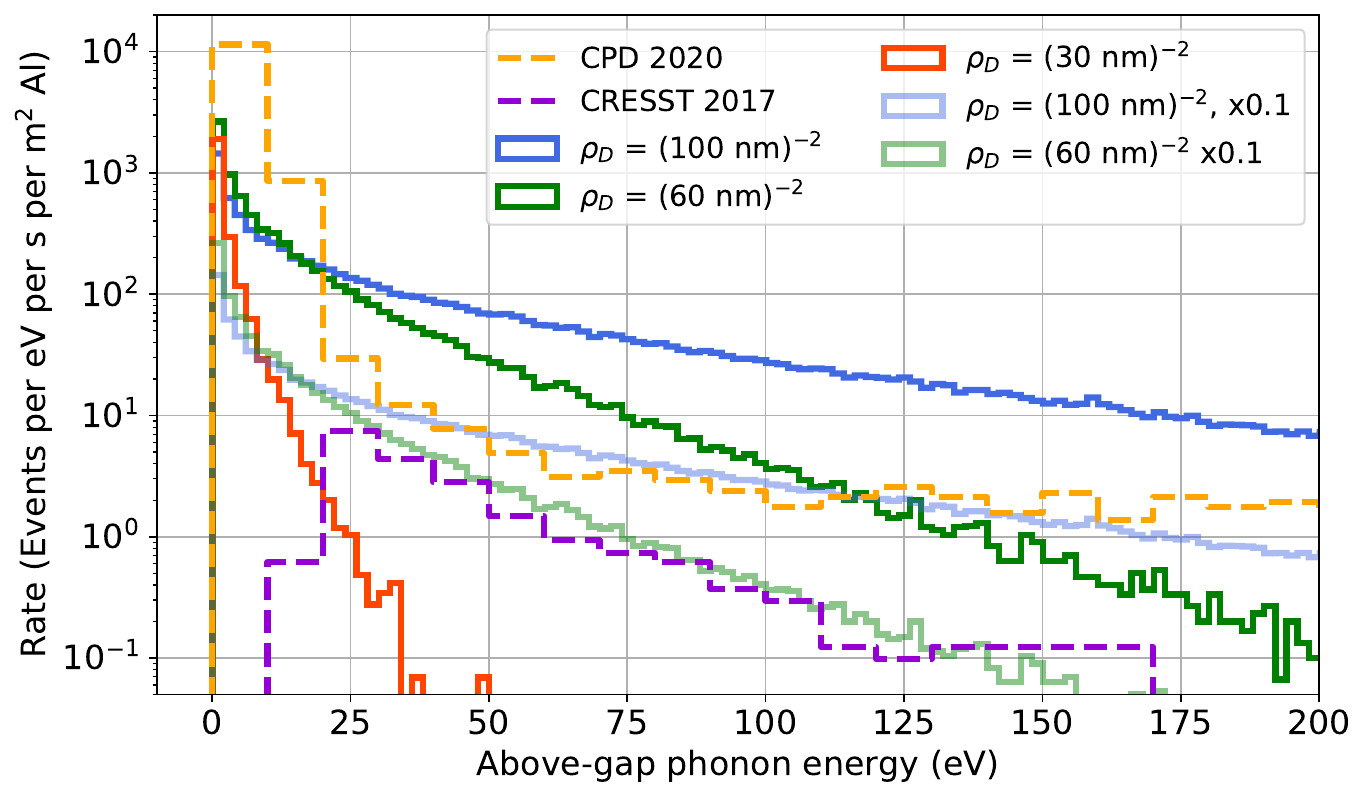}
\caption{\label{fig:simulated_spectrum} The simulated spectrum of phonon energies for dislocation relaxation events at various dislocation densities and at a bulk stress $\sigma_0 = $ 120 MPa for a simulated detector 1 week after cooldown (see text). Experimental measurements from Refs. \cite{CRESST2017, CPDV1} are superimposed. The lighter colored blue and green lines correspond to the dark blue ($\rho_D = $ (100 nm)$^{-2}$) and dark green ($\rho_D = $ (60 nm)$^{-2}$) spectrum scaled down by one order of magnitude, to account for systematic uncertainties (see text). See Sec. \ref{section:spectrum_sim} for details on the simulation and Appendix \ref{appendix:detector_scaling} for details on the scaling of experimental results.} 
\end{figure}

After drawing values for $h_{\mathrm{max}}$ and $\tau$, the total above-gap phonon energy emitted into the substrate $E_{\mathrm{tot}}$ was calculated for each simulated event (see Fig. \ref{fig:simulated_spectrum}).

The total number of active pinning sites was estimated by assuming that every (otherwise unimpeded) dislocation travels a distance $d = \epsilon /(\rho_D b)$ where $\epsilon \approx 4 \times 10^{-3}$ is the relative strain created by differential thermal contraction, and by assuming that the mean free path for the dislocation is $\lambda = 1/(\sqrt{\rho_D})$.

For this depinning process, I only consider one type of lock, specifically the Lomer-Cottrell lock. I define $\phi$ as the fraction of dislocation interactions which will result in Lomer-Cottrell pinning, and take $\phi = 24/144$, \cite{LockStrengths} assuming that the interacting dislocations Burgers vector are randomly chosen from the 144 independent Burgers vectors configurations for interacting dislocations traveling on two separate slip plane on the Thompson tetrahedron \cite{DislocationInteractionsThompson, HullBacon, LockStrengths}. Only 24 configurations will result in a Lomer-Cottrell lock \cite{LockStrengths, DislocationInteractionsThompson, HullBacon}, the pinning site I consider in this paper. (Most of the remainder of interactions produce other types of locks, e.g. Hirth or Lomer locks, as well as interactions that are ``glissile,'' i.e. more weakly interacting. \cite{LockStrengths}) The fraction $f$ of appropriately pinned dislocations is, therefore, roughly
\begin{eqnarray}
    f \approx 1 - e^{-d/ \lambda} = 1 - \exp \Big( - \frac{\epsilon \phi }{b \sqrt{\rho_D}} \Big)
\end{eqnarray}
which gives a total event rate
\begin{eqnarray}
    R(t) = \frac{A_{\mathrm{Al}} \rho_D}{\Gamma t} \bigg( 1 - \exp \Big( - \frac{\epsilon \phi }{b \sqrt{\rho_D}} \Big) \bigg)
\end{eqnarray}
using Eq. \ref{equation:rate}. One week after cooldown, the total rate of events that emit non-zero amounts of above-gap energy is, therefore, around 6.0, 16.2, and 16.4 mHz/(Al mm$^2$) for $\rho_D$ = (30 nm)$^{-2}$, (60 nm)$^{-2}$, and (100 nm)$^{-2}$ respectively.

Figure \ref{fig:simulated_spectrum} shows the result of these simulations. At values of $\rho_D$ derived from literature expectations \cite{DislocationDensity1, AlInternalFriction, DislocationDensity3}, the simulated spectral shapes broadly match experimental observations, but seem to overestimate the observed rate by about an order of magnitude. As in this model the spectral rate is correlated to the spectral shape, choosing different values of $\rho_D$ cannot simultaneously fit the rate and shape of the experimental spectra. I instead show that an ad-hod scaling down of the spectra by an order of magnitude (see Fig. \ref{fig:simulated_spectrum}) produces better agreement with the data.

Two sources of uncertainty might explain this discrepancy between simulation and experiment. First, as discussed in Appendix \ref{appendix:detector_scaling}, uncertainty in the precise time at which a published spectrum was taken and the aluminum coverage of a given detector might introduce a scaling of the experimental curves by as much as a factor of two. More importantly, in this model the total estimated spectra rate scales as 
\begin{eqnarray}
    \frac{dR}{dE}(t) \propto \frac{1}{\Gamma} \propto \frac{1}{\mathbb{A} \sqrt{M W_0}}
\end{eqnarray}
Given that $\mathbb{A}, M, W_0$ were all only estimated at the $\mathcal{O}(1)$ level by Mott \cite{MottTunneling}, it is easy to imagine that the true value of $\Gamma$ might differ from my estimate by an order of magnitude, explaining the discrepancy between these simulations and previous experiments. Finally, other considerations (e.g. spatially varying dislocation densities and dislocation-grain boundary interactions) may introduce additional corrections to this model in higher fidelity simulations.

\section{Experimental Evidence}

In addition to the well documented observation of low energy excesses in dark matter direct detection experiments, several pieces of experimental evidence support this model.

Low temperature relaxation of stressed aluminum through non-thermal processes (i.e. tunneling) is supported by the literature. Aluminum has been observed to deform below liquid helium temperatures \cite{LowTempAlCreep} and over very long periods of time \cite{TAMULowTAlCreep}. This long period deformation has been reported to occur in individual sudden relaxation events interspersed with long periods where no relaxation occurs \cite{JumplikeDeformation}, possibly corresponding to individual dislocation relaxation events. In addition to these ``quantum creep'' type observations, dislocation tunneling through pinning sites has been observed in ``internal friction'' type experiments, which have allowed for the observation of dislocation pinning and depinning well below the superconducting transition \cite{AlInternalFriction, PinningTunneling, DislocationTunneling}. Work hardening type effects (i.e. pinning increasing with increasing dislocation densities) have also been observed \cite{AlInternalFriction}, strongly suggesting that dislocation-dislocation interactions play a role in governing thin film aluminum relaxation. At higher temperatures, thin aluminum films on silicon have been observed to yield under thermal stress \cite{ThinAlFilmDeformation}.

The interaction of dislocations with phonons is well established through theory and simulation (see e.g. Refs. \cite{Dislon} and \cite{ ThermalConductivityDislocation}). Unfortunately, the regime described in this paper (the sudden deceleration of dislocations creating high energy phonons) does not appear to be well-studied.

\section{Implications for Detector Design}

Events of these types would form both backgrounds and noise (sub-threshold) events for current generation low temperature detectors that use aluminum thin films. Reducing the rate of these excess events is, therefore, key to performing low background measurements and building detectors with good energy resolution.

Most obviously, the rate of these events could be reduced by reducing the amount of aluminum on these devices. Of course, excess non-instrumented metal films on the device surface also serve to ``soak up'' phonons, reducing the energy absorbed in individual sensors from phonon mediated events.

Changes to the material properties of the film could also reduce the impacts of the LEE. As suggested by Fig. \ref{fig:simulated_spectrum}, increasing the dislocation density (through e.g. work hardening) may best serve to decrease the rate of high energy events. Decreasing interface grain sizes to reduce the number of dislocations that are long enough to spontaneously relax could have a similar effect. The opposite approach (annealing to decrease dislocation densities) seems likely to produce limited improvements, given the simulations shown in Fig. \ref{fig:simulated_spectrum} and that dislocations will naturally be seeded during the cooldown \cite{ThinFilmSim}, effectively undoing some of the effects of annealing. Any changes to the material properties of the aluminum films would need to preserve both a good coupling to the crystal phonon system and a high quasiparticle diffusion length, potentially posing difficulties when e.g. increasing dislocation density or decreasing interface grain sizes.

Finally, attempts could be made to artificially depin these metastable dislocations by temporarily \textit{increasing} the stress in the aluminum films on the device surface. This could be achieved by e.g. flexing the detector substrate (which would be impractical for thick dark matter detectors) or by compressing the films from above to increase the von Misses deviatoric stress in the film. Once the excess film stress was reduced, the films would be left in a more relaxed state, presumably with fewer dislocation relaxation events.

If eliminating relaxation events from aluminum proves difficult, substituting other materials for aluminum (see Appendix \ref{appendix:other_materials}) in low temperature calorimeters may serve to lower the rate of LEE events. Other materials used for e.g. phonon collection fins would need to have similar superconducting transition temperatures (to retain sensitivity to similar parts of the phonon spectrum), to have good quasiparticle propagation, and efficient phonon absorption from the substrate. Finally, detector schemes that use Josephson junctions (e.g. qubit-based detectors \cite{QCD, SQUATIdeaPaper}) may struggle to completely eliminate aluminum in their designs, given the favorable properties of aluminum for constructing Josephson junctions.

\section{Tension with Experiments}

This paper aims to present a proof-of-concept model illustrating that stress relaxation in low temperature aluminum can create phonon bursts consistent with the LEE, making simplifying assumptions to more easily model the core relaxation behavior. Two areas where the model is in tension with experimental results should be explicitly noted.

First, this model predicts that $\sim$ 10-20 \% of phonons from a dislocation relaxation event are emitted into the film (as opposed to into the crystal). Multiple experiments including Ref. \cite{CRESSTSingles} have observed that events that primarily couple to the phonon system seem to entirely couple to the phonon system. While more accurate simulation and experimental effects are likely to significantly relax this tension (see Section \ref{section:phonon_creation}), future experimental and phenomenological work should focus on this potential discrepancy.

Second, the simulated spectra in Fig. \ref{fig:simulated_spectrum} differ in rate from the experimental results \cite{CPDV1, CRESST2017}. As discussed in section \ref{section:spectrum_sim}, this discrepancy can be resolved by assuming a somewhat different value of $\Gamma$ than estimated by Mott \cite{MottTunneling}.

\section{Conclusion}

In this paper, I employ a model to show that the relaxation of thermal stress in aluminum films (as mediated by dislocations) can produce sudden bursts of phonons broadly consistent with the LEE. This model closely reproduces the observed rate of LEE events, the qualitative rate time dependence, the approximate spectrum of event energies, and the primary coupling to the phonon system observed for LEE events. Of course, any given experiment might host multiple types of LEE (e.g. experiments with a significant amount of dielectrics might also observe the events described in Ref. \cite{EssigTrackInduced}), meaning that this model might not fully describe LEE observations in any given experiment.

Further experiments (e.g. testing the relation between excess rates and aluminum coverage) as well as phenomenological work (e.g. atomistic modeling of dislocation depinning through tunneling, the radiation of phonons during deceleration, and the effects of realistic dislocation/grain configurations at the film-crystal interface, see Appendix \ref{sec:future_work}) are needed to further support or contradict this model. In light of this work, relaxation processes in other materials commonly used in low temperature detectors (e.g. tungsten, niobium and gold) should be more fully considered, as they may be the dominant source of the current excess, or may be responsible for events observed in future experiments.

Recent interest in the links between the low energy excess and the anomalous decoherence of superconducting qubits \cite{StressBackgroundsPaper} suggests that extensions of this model may also provide an explanation for the so-called ``quasiparticle poisoning'' \cite{Serniak2018}. Low energy events which are predominantly absorbed locally in films (rather than mediated through crystal phonons) would be an especially important for qubit systems.

\begin{acknowledgments}
I wish to thank Daniel Ega\~{n}a-Ugrinovic, Matt Pyle, Daniel McKinsey, Rouven Essig, Marco Costa, Michael Williams, Daniel Baxter, Noah Kurinsky and Junwu Huang for helpful conversations.

This research was supported by US Department of Energy Grant DE-SC0022354, and in part by the Perimeter Institute for Theoretical Physics. Research at the Perimeter institute is supported through the Government of Canada through the Department of Innovation, Science and Economic Development, and by the Province of Ontario through the Ministry of Colleges and Universities.
\end{acknowledgments}

\begin{appendices}
\counterwithin*{equation}{section}
\renewcommand\theequation{\thesection\arabic{equation}}

\section{Appendix: Dislocation Motion Numerical Simulation}
\label{appendix:simulation}

As described in the overview, this paper relies on two simple Python-based simulations \cite{SimulationGithub}. 

\subsubsection{Dislocation Motion and Phonon Radiation Simulation}

To simulate the motion and phonon radiation from a short length $dL$ of dislocation impacting the film-crystal interface, the following analytic equations of dislocation motion (also described in section \ref{section:phonon_creation}) were numerically solved using SciPy's ``solve\_ivp'' method. Following Refs. \cite{HirthDislocations} and \cite{IntefaceForce}, I set
\begin{eqnarray}
    \omega(x) = \frac{c_t}{\gamma b} \exp \bigg( \frac{1}{2} \mathrm{W}\Big( -4 \gamma^2 \frac{\mu_s - \mu_f}{\mu_s + \mu_f} \frac{b^2}{x^2}  \Big) \bigg) \\
    m^*(x) = \frac{dm}{dL}(x) = \frac{\mu_f b^2}{4 \pi c_t^2} \ln \bigg( \frac{c_t}{\gamma b \omega(x)} \bigg) \\
    F_\mathrm{rad}^* (x, v) = \frac{dF_{\mathrm{rad}}}{dL} (x, v) = - \frac{\mu_f b^2 v \omega(x)}{8 c_t^2} \\
    F^*_\mathrm{interface}(x) = \frac{dF_{\mathrm{interface}}}{dL} (x) = \frac{\mu_f b^2}{4 \pi} \frac{\mu_s - \mu_f}{\mu_s + \mu_f} \frac{1}{x}\\
    F^*_\mathrm{tot}(x, v) =  F_\mathrm{rad}^* (x, v) + F_\mathrm{interface}^* (x) + \tau b\\
    \frac{dx^2}{dt^2} = \frac{F_\mathrm{tot}^*(x, v)}{m^*(x)}
\end{eqnarray}
where W is the Lambert W function and other symbols are defined in the main text. The initial conditions were $x(0) = h$ and $dx/dt(0) = 0$, and I took $b=0.286$ nm, $c_t = 3040$ m/s, $\mu_f = 25$ GPa, and $\mu_s$ = 51 GPa.

By numerically solving these differential equations, $x(t), v(t), \omega(t)...$ were obtained (see Fig. \ref{fig:dislocation_motion}). The power radiated into phonons by a unit length of dislocation at each moment in time could also be computed using
\begin{eqnarray}
    P^*_\mathrm{rad}(t) = F^*_\mathrm{rad}(t) v(t) - \frac{1}{2} \frac{dm}{dt}(t) v^2(t)
\end{eqnarray}
assuming that the reduction in effective dislocation mass occurs through the radiation of phonons as described in the main text and discussed below.

During the deceleration process, phonons will be emitted in one of two directions: either into the crystal through the crystal-film interface (when the dislocation is damped while moving toward the interface, i.e. $F^*_\mathrm{rad} > 0$, i.e. $v < 0$) or into the film (when the dislocation is damped while moving away from the interface, i.e. $F^*_\mathrm{rad} < 0$, i.e. $v > 0$). The total amount of energy radiated in each direction can be computed by summing
\begin{eqnarray}
    \frac{dE_\mathrm{crystal}}{dL} = E^*_\mathrm{crystal} = \sum_{v(t) < 0} P^*_\mathrm{rad}(t) \Delta t \\
    \frac{dE_\mathrm{film}}{dL} = E^*_\mathrm{film} = \sum_{v(t) > 0} P^*_\mathrm{rad}(t) \Delta t
\end{eqnarray}
where $\Delta t$ is the time step of the simulation. Similarly, the energy of phonons emitted at a given time (and, therefore, the spectrum of emitted phonons) can be calculated using $E_\mathrm{phonon}(t) = \hbar \omega(t)$, with the sign of $v(t)$ used to determine if these phonons are emitted into the film or crystal. From this relation, the energy of e.g. above-superconducting-gap phonons emitted into the crystal can be calculated using
\begin{eqnarray}
    E^*_\mathrm{crystal, above-gap} = \sum_{v(t) < 0, \hbar \omega(t) > 2 \Delta_{Al}} P^*_\mathrm{rad}(t) \Delta t
\end{eqnarray}
where $\Delta_{Al}$ is the superconducting bandgap energy.

By numerically solving this differential equation and performing the above sums, $E^*(h, \tau)$ can be simulated for a grid of realistic $h, \tau$ points.

As noted in the main text, this approach is only a one-dimensional analytic approximation of the true dynamics of the system. See Appendix \ref{sec:future_work} for more details on how these simulations could be more realistically performed in future work.

Two caveats of this approach warrant explicit recognition. First, as modeled in Sec. \ref{section:phonon_creation}, the dislocation effective mass changes as a function of position above the interface. While decreases in the dislocation kinetic energy through a decrease in effective mass are easily interpreted as the radiation of kinetic energy through the emission of phonons, \textit{increases} in kinetic energy due to an increase in mass are more difficult to interpret.

Second, no ``elastorelativistic'' corrections are applied in this model, meaning that the dislocation can accelerate beyond $c_t$ for high stresses and large initial distances between the dislocation and interface. While this system is somewhat unique in that e.g. dislocation-phonon damping is highly suppressed, and while supersonic dislocation motion has been experimentally observed \cite{SupersonicDislocationDiamond}, supersonic dislocations remain controversial. Future work should attempt to more accurately model the motion of the dislocation close to $c_t$.

\subsubsection{Monte-Carlo Based Spectrum Simulation}

After simulating the emitted phonon energy per unit length of dislocation $E^*(h, \tau)$, I calculated the total energy emitted by the entire dislocation by integrating
\begin{eqnarray}
    E_{\mathrm{tot}} = \int_y E^*(h(y), \tau) dy \\
     = \int_0^{h_{\mathrm{max}}} E^*(h(y), \tau) \frac{dy}{dh} dh
\end{eqnarray}

In general, the shape $h(y)$ of the pinned dislocation is governed by self-interaction and interactions with the pinning site, shear stress field, and the interface, leading to a complex shape likely only determinable using simulation (see Appendix \ref{sec:future_work}). Here, I make the assumption that the dislocation shape is linear between the pinning site and the interface, meaning that for a dislocation with total width $L(\tau)$ (see Eqn. \ref{eqn:stress_length_relationship}),
\begin{eqnarray}
    \frac{dh}{dy} = \frac{h_{\mathrm{max}}}{L(\tau)}
\end{eqnarray}
Using this assumption and the simulated $E^*$, $E_{\mathrm{tot}}(h_\mathrm{max}, \tau)$ can be calculated, as shown in Fig. \ref{fig:dislocation_energy_2d}.

Changes to this model of the initial dislocation shape $h(x)$ (e.g. assuming that the dislocation ends are pinned away from the interface, rather than at the interface) may increase the fraction of event energy emitted into the crystal phonon system, potentially resolving some of the tension discussed in section \ref{section:phonon_creation}.

This calculation yields a total energy $E_\mathrm{tot}(h_\mathrm{max}, \tau)$ of e.g. phonons emitted into the crystal above the aluminum superconducting bandgap for a given resolved shear stress $\tau$ and pinning distance $h_\mathrm{max}$. Using a Monte Carlo method, these were used to simulate a spectrum of events expected in a given device (see section \ref{section:spectrum_sim}). $h_\mathrm{max}$ and $\tau$ were drawn from
\begin{eqnarray}
    \mathrm{Pr}(h_{\mathrm{max}})_\tau = \rho_D L(\tau) e^{-h_{\mathrm{max}} / (\rho_D L(\tau))} \\
    \mathrm{Pr}(\theta) = \cos (\theta) \\
    \tau(\theta) = \frac{\sigma_0}{2} \sin (2 \theta)
\end{eqnarray}
assuming $\sigma_0 = $ 120 MPa, and for a variety of realistic $\rho_D$ (see section \ref{section:spectrum_sim}). From these drawn values, $E_\mathrm{tot}(h_\mathrm{max}, \tau)$ was calculated and converted into a spectrum (see Fig. \ref{fig:simulated_spectrum}). These spectra were scaled assuming a total rate of events given by Eqn. \ref{equation:rate} (see section \ref{section:spectrum_sim}).

\section{Appendix: Scaling Detector Spectra}
\label{appendix:detector_scaling}

Comparing simulated results to real data requires scaling to allow for an equal-footed test. In this appendix, I describe how the data in Refs. \cite{CPDV1, CRESST2017} (with data taken from the public repository described in Ref. \cite{EXCESSReview}) were scaled for Fig. \ref{fig:simulated_spectrum}. In general, the necessary information (particularly how long the detector was cold when the measurement was performed and the amount of aluminum film on the device surface) is not published, requiring some parameters to be estimated. Better understanding these parameters may resolve some of the tension between simulation and experimental results seen in Fig. \ref{fig:simulated_spectrum}.

For the CRESST-III main detector described in Ref. \cite{CRESST2017}, the detector was assumed to have been cold for 1 year, and the aluminum fins were assumed to be the same size as the aluminum fins for the CRESST-III light detector (LD) \cite{CRESSTDetectorTechnical}.

For the CPD detector in Ref. \cite{CPDV1} the detector was assumed to have been cold for 6 weeks, and to have 2\% of its 45.6 cm$^2$ surface area covered by aluminum \cite{CPDTechnical}.

To compensate for the amount of time detectors were cold, the spectra given in Ref. \cite{EXCESSReview} were scaled up by a factor of (Time Cold)/(1 Week).

\section{Appendix: Other Materials}
\label{appendix:other_materials}

This model relies upon free (unpinned) dislocations being able to easily move around in the crystal lattice under thermally induced stresses, i.e. that the Peierls stress is low compared to the thermally induced stress. In the case of aluminum, this assumption holds (see Sec. \ref{section:film_stress_deformation}).

Copper, gold, and other FCC metals commonly used in low temperature devices have similarly low Peierls stresses, and might, therefore, create phonon bursts similar to those created by aluminum when relaxing at low temperatures. Key differences between these materials and aluminum include the effective dislocation mass and the Debye frequency (which would affect the depinning rate), as well as the speed of sound, Burgers vector length and shear modulus (which effect the phonon emission process). The strength of dislocation pinning to either impurities or forest dislocations would also be expected to vary, which would in turn effect the depinning rate and the scale of depinned dislocations. Finally, typical dislocation densities in these materials may differ from those assumed here for aluminum. The extension of this model to other materials is left for future work. BCC metals with somewhat low Peierls stresses such as niobium \cite{NbPeierlsStress} may also behave similarly to aluminum.

Materials with higher Peierls stresses such as tungsten would not be expected to deform through the mechanisms described in this paper, as their Peierls stresses greatly exceed thermally induced stresses \cite{NbPeierlsStress}, making dislocation motion through the crystal lattice difficult. While these materials might yield after thermal stressing (through e.g. cracking or interface delamination, see Ref. \cite{WDelamSim} for the delamination of W on Si), they would not be expected to deform through the motion of dislocations as described in this model. Experimentally, thin-film tungsten has been deposited with stresses varying by several GPa (see e.g. Refs. \cite{WMeasuredStress, WMeasuredStress2}), suggesting that thermally induced 100 MPa-scale stress changes are not likely to cause macroscopic yielding (as has been observed for aluminum \cite{ThinAlFilmDeformation}). 

\section{Appendix: Future Work}
\label{sec:future_work}

The model presented in this work could be improved by applying more existing simulation codes. In this section, I suggest areas for future improvement, ordered similarly to the sections in the main body of this paper.

Detailed predictions of the tunneling rate (that go beyond the simple estimates of Mott \cite{MottTunneling} used in this work) would need to rely on more accurate knowledge of the dislocation-dislocation interaction potential (assuming that these pinning sites are responsible for observations). This could be calculated as a function of resolved stress, dislocation-dislocation orientation, and dislocation distance using a density functional theory-based package such as VASP \cite{VASP}, similar to the work done for impurity pinning sites \cite{AlMgImpurities}. Once these potentials are known, tunneling rates for each configuration could be calculated.

To more accurately calculate the energy emitted by a relaxation event, a number of additional pieces of information are needed. First,  knowledge of the distribution of grain sizes at the crystal-film interface is needed. Likely, this will be provided both from characterization of real films (using e.g. Transmission Electron Microscopy) but also simulations of film deposition (see, e.g., Refs. \cite{GrainGrowth2D} and \cite{GrainGrowth3D}).

From this knowledge of the structure of the film, the dislocation seeding and stressing process during cooldown can be simulated in three dimensions. This should likely be undertaken with a discrete dislocation dynamics package such as ParaDIS \cite{ParaDIS} which would include e.g. boundary conditions that reflect thin films and realistic grain sizes. This simulation would inform
\begin{itemize}
    \item The density and stress distribution of locked dislocations
    \item The shape of the pinned dislocation (which informs the amount of stored energy and the speed of the unpinned dislocation when impacting the interface)
    \item The length of the dislocation that will impact the film-crystal interface (rather than a grain boundary)
\end{itemize}
Course grained atomistic approaches (see Ref. \cite{CourseGrained}) could also be employed. The involvement of experts would of course be key to selecting the most appropriate simulation technique.

Finally, the phonon-radiating dislocation impact at the film-crystal interface could be more accurately simulated using existing codes. Although simulations of dislocation-phonon interactions based on a course grained atomistic approach have been performed (see e.g. Ref. \cite{PhononDislocationSim}), no existing tools seem to be designed to simulate the radiation of high energy phonons by decelerating dislocations. Some combination of existing codes and extensions based on the theory of dislocation-phonon interactions (perhaps using the ``dislon'' approach \cite{Dislon}) should be employed to more accurately simulate the spectrum and direction of phonons radiated during this deceleration process. This simulation could also be applied to the impact of dislocations on e.g. grain boundaries or forest dislocations. Impacts with grain boundaries or forest dislocations are very difficult to imagine understanding analytically, and could give insight into a range of local effects (as described in e.g. Ref. \cite{CRESSTSingles}).

This three-dimensional simulation of phonon-radiation would start with the dislocation in a just-unpinned state, track its acceleration toward the grain boundary under the resolved shear stress field (naturally including ``elastorelativistic'' effects), and simulate the radiation of phonons from the dislocation as it decelerates at the interface. From this, a spectrum of emitted phonons for a given relaxation event could be calculated. Combined with knowledge of the film configuration near the interface, this could be used to create a Monte Carlo type simulation to generate spectra, which could be compared to LEE spectra measured in real detectors.

\end{appendices}


\bibliography{apssamp}

\end{document}